\documentstyle[twoside,fleqn,espcrc2,epsf]{article}

\title{
Scalar Quarkonium and the Scalar Glueball}

\author{Don Weingarten\\
IBM Research, P.O.~Box 218,
Yorktown Heights, NY 10598\\}

\begin{document}

\begin{abstract}

Valence approximation glueball mass and decay calculations support the
identification of $f_J(1710)$ as the lightest scalar glueball. An
alternate glueball candidate is $f_0(1500)$.  I present evidence for
the identification of $f_0(1500)$ as $s\overline{s}$ quarkonium.
 
\end{abstract}

\maketitle

\section{Introduction}

The identification of $f_J(1710)$ as the lightest scalar glueball is
supported at present by two different sets of calculations.

For the valence approximation to the infinite volume continuum limit of
the lightest scalar glueball mass, a calculation on
GF11~\cite{Vaccarino}, using 25000 to 30000 gauge configurations, gives
$1740 \pm 71$ MeV. An independent calculation by the
UKQCD-Wuppertal~\cite{Livertal} collaboration, using 1000 to 3000 gauge
configurations, gives $1625 \pm 94$ MeV when extrapolated to zero
lattice spacing according to Ref.~\cite{Weingarten94}.  The GF11 and
UKQCD-Wuppertal data combined predicted $1707 \pm 64$ MeV.  The
calculation with larger statistics and the combined result both favor
$f_J(1710)$ as the lightest scalar glueball.

The mass calculations by themselves, however, leave open the possibility
that the lightest glueball may have too large a total width to be found
in experiment. A valence approximation calculation on
GF11~\cite{Sexton95} of couplings for glueball decay to all possible
pseudoscalar pairs, $\pi + \pi$, $K + \overline{K}$, $\eta + \eta$, and
$\eta + \eta'$, using a $16^3 \times 24$ lattice at $\beta = 5.7$, gets
$108 \pm 29$ MeV for the total two-body width. Based on this number, any
reasonable guess for the width to multibody states yields a total width
small enough for the lightest scalar glueball to be seen easily in
experiment. In fact, the predicted total two-body width agrees with the
$f_J(1710)$ width of $99 \pm 15$ MeV of Ref.~\cite{Long}, as do the
partial widths to individual channels.

Among established resonances with the quantum numbers to be a scalar
glueball, aside from $f_J(1710)$ all are clearly inconsistent with the
mass calculation expect $f_0(1500)$. The mass of $f_0(1500)$ is still
more than 3 sigma away both from the prediction with larger statistics
or from the combined result. Ref.~\cite{Sexton95} proposes to interpret
$f_0(1500)$ as dominantly composed of $s\overline{s}$ scalar quarkonium.
The interpretation of $f_0(1500)$ as $s\overline{s}$ quarkonium,
however, encounters three difficulties.  First, it appears possible that
the gap between $1740 \pm 71$ MeV and 1500 MeV might simply be an error
arising from the valence approximation.  Second, $f_0(1500)$ does not
seem to decay mainly into states containing an $s$ and an $\overline{s}$
quark~\cite{Amsler1}.  Third, the Hamiltonian of full QCD couples
quarkonium and glueballs so that $f_J(1710)$ and $f_0(1500)$ could both
be linear combinations of quarkonium and a glueball, perhaps even half
glueball and half quarkonium each.  

In the remainder of this article, I show that the pattern of established
quarkonium masses, an estimate of the error in valence approximation
mass calculations, a calculation of the $s\overline{s}$ scalar
quarkonium mass and a model of quarkonium-glueball mixing help resolve
these difficulties and support the interpretation of $f_J(1710)$ as
dominantly a glueball and $f_0(1500)$ as dominantly $s\overline{s}$
scalar quarkonium.

\section{Comparison with Experiment}

The simplest piece of evidence suggesting that $f_J(1710)$ is mainly a
glueball while $f_0(1500)$ is mainly $s\overline{s}$ quarkonium is
provided by the pattern of masses among established meson resonances.

\begin{figure}
\epsfxsize=63mm
\epsfbox{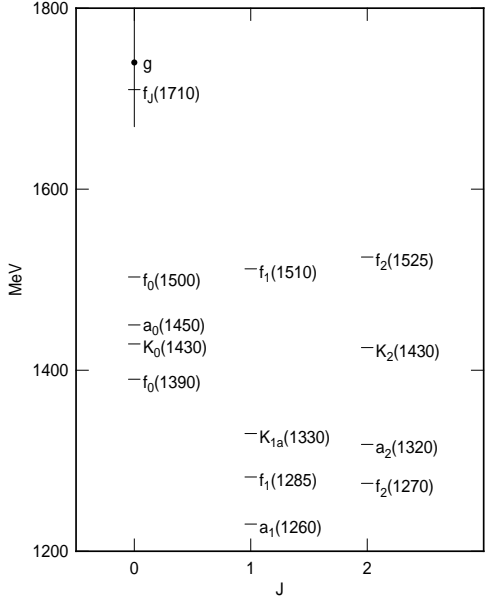}
\vskip -9mm
\caption{Established $0^{++}$, $1^{++}$ and
$2^{++}$ resonances and their strange $0^+$, $1^+$ and $2^+$ 
partners.}
\vskip -6mm
\label{fig:spect}
\end{figure}

Figure~\ref{fig:spect} shows the established $0^{++}$, $1^{++}$ and
$2^{++}$ resonances and their strange $0^+$, $1^+$ and $2^+$ partners,
with the omission only of $f_0(980)$, $a_0(980)$ and $f_1(1420)$, all
irrelevant to glueball spectroscopy.  The scalar glueball
prediction of $1740 \pm 71$ MeV is labeled ``g''.  If $f_J(1710)$ were
also omitted, the diagram would include exactly one state for each
possible combination of $u$, $d$ and $s$ quark-antiquark pairs with
orbital angular momentum 1 and total spin angular momentum 1.  For all
three values of total angular momentum, the $a_J$ states are 
isovector combinations of $u$ and $d$ quarks. From their masses and from
their decay modes, it is not hard to show that the lower $f_J$ states
are dominantly isoscalar combinations of $u$ and $d$ quarks.  The $K_J$
states are all combinations of one $s$ and a $u$ or $d$.  The higher
$f_1$ and $f_2$ states both decay dominantly into states including both
an $s$ and an $\overline{s}$. Thus the higher $f_1$ and $f_2$ are
primarily $s\overline{s}$.

What about $f_J(1710)$ and $f_0(1500)$?  From the discussion of the
other states in the picture, it is clear that $f_0(1500)$ is about where
it should for an $s\overline{s}$ quark-antiquark meson. The mass of
$f_J(1710)$, meanwhile, is far from the region expected for
$s\overline{s}$ quarkonium but fits the predicted glueball mass.

\section{Valence Approximation Errors}

Can the disagreement between $1740 \pm 71$ MeV and the mass of
$f_0(1500)$ be an error arising from the valence approximation?  In the
GF11 calculation of eight infinite volume continuum limit hadron
masses~\cite{Butler}, the largest disagreement with experiment was
6\%. Out of eight random variables, one is expected to be above its mean
by a standard deviation or more. So 6\% is a plausible estimate for the
one sigma upper bound on valence approximation errors for light hadron
masses.  A 6\% valence approximation error bound on the glueball mass
would be 100 MeV.  

A simple argument suggests, however, that as in the case of meson decay
constants, full QCD is likely to yield a predicted value higher than the
valence approximation prediction and will thus agree with 1500 MeV no
better than does the valence approximation result.  The valence
approximation may be viewed as replacing the momentum and frequency
dependent color dielectric constant arising from quark-antiquark vacuum
polarization with its low momentum limit.  At low momentum, then, the
effective color charge appearing in the valence approximation will agree
with the low momentum effective charge of the full theory.  The valence
approximation's effective color charge at higher momentum can be
obtained from the low momentum charge by the Callan-Symanzik equation.
As a consequence of the absence of dynamical quark-antiquark vacuum
polarization, the color charge in the valence approximation will fall
faster with momentum than it does in the full theory, and therefore be
smaller than it should be at high momentum.  Since the scalar glueball
is significantly heavier and significantly smaller in radius than the
$\rho$, whose mass is used in the glueball mass calculations to set the
color charge at low momentum, the glueball should include more high
momentum chromoelectric field than built into the determination of the
color charge.  This high momentum part of the scalar glueball sees a
smaller color charge in the valence approximation than it does in full
QCD and therefore contributes less than it should to the glueball's
total energy. Thus the valence approximation scalar glueball mass will
lie below the full QCD prediction.

In passing, it is perhaps also useful to consider how far the
valence approximation, finite lattice spacing decay couplings are likely
to be from the real world. From the comparison of finite lattice spacing
valence approximation hadron masses with their values in the real world,
I would expect an error of 15\% or less in going to the continuum limit
and another 6\% or less arising from the valence approximation.  The
total predicted width for glueball decay to two pseudoscalars should
then have an error of less than 50\%. A 50\% increase in our predicted
two-body decay width, combined with any reasonable corresponding guess
for multibody decays, gives a total glueball width small enough for the
particle to be observed easily.

\section{Scalar Quarkonium Mass Calculation}

\begin{figure}
\epsfxsize=65mm
\epsfbox{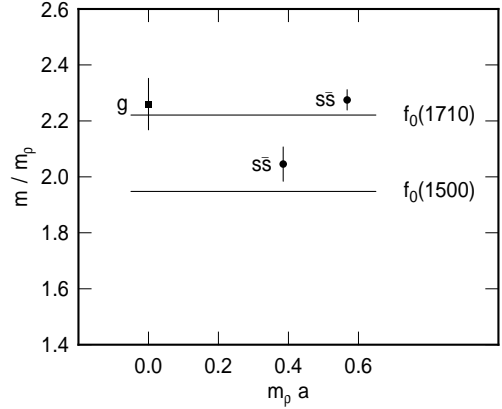}
\vskip -9mm
\caption{Scalar $s\overline{s}$ quarkonium masses for two different
values of lattice spacing.}
\vskip -9mm
\label{fig:bothbetas}
\end{figure}

Weonjong Lee and I~\cite{Lee} are in the process of calculating, in the
valence approximation, the mass of scalar quark-antiquark bound states
as an additional test of the hypothesis that $f_0(1500)$ is mainly
$s\overline{s}$ quarkonium.  Figure~\ref{fig:bothbetas} shows the scalar
$s\overline{s}$ mass we have obtained at two different values of lattice
spacing.  The lattice period in both cases is nearly 2.3 fm.  The square
at zero-lattice-spacing is the continuum limit of the scalar glueball
mass, and horizontal lines show the $f_0(1500)$ and $f_0(1710)$ masses,
all in units of the $\rho$ mass. It is clear from the two points so far
that the valence approximation $s\overline{s}$ mass in the continuum
limit with lattice period fixed at 2.3 fm will lie significantly below
1710 MeV and significantly below the predicted scalar glueball mass. It
is shown in Ref.~\cite{Lee} that taking the infinite volume limit of the
$s\overline{s}$ mass will not change this conclusion.  These results
tend to support the interpretation of $f_0(1500)$ as an $s\overline{s}$
state, and tend to exclude the possibility that $f_J(1710)$ might be an
$s\overline{s}$.

\section{Glueball-Quarkonium Mixing Model}

I will now consider a simple model of the mixing between the scalar
glueball and isosinglet quarkonium states. The result will be that the
the mixed glueball is more than 75\% pure glueball and the mixed
quarkonium states are more than 75\% pure quarkonium. The mixing is still
sufficient, however, to give a strong suppression of $K\overline{K}$
decays from the $f_0(1500)$. An orthogonal model of glueball-quarkonium
mixing, which takes $f_0(1500)$ to be primarily a glueball, is discussed
in Ref.~\cite{Amsler2}.

To leading order in the valence approximation, with valence quark
annihilation turned off, corresponding isotriplet and isosinglet states
composed of $u$ and $d$ quarks will be degenerate. For the scalar meson
multiplet, the isotriplet state of $u$ and $d$ quarks has a mass of 1450
MeV~\cite{Amsler3}. An isosinglet mass of 1390 MeV is reported by the
Crystal Barrel collaboration~\cite{Amsler1}. Mark III finds 1430
MeV~\cite{MarkIII}. The isosinglet-isovector splitting gives a measure
of the strength of the valence quark annihilation process for isosinglet
scalar mesons. Some part of this splitting will arise from annihilation
into a scalar glueball. If we assume the splitting arises entirely from
coupling to the scalar glueball and take the lower isosinglet mass, 1390
MeV, we should get an upper estimate on the strength of this coupling.
Since scalar meson masses appear to depend relatively weakly on quark
mass, let us introduce the further assumption that this valence quark
annihilation amplitude is the same for $u\overline{u}$, $d\overline{d}$
and $s\overline{s}$.

The structure of the Hamiltonian coupling together the scalar glueball,
the scalar $s\overline{s}$ and the scalar $(u\overline{u} +
d\overline{d})$ isosinglet becomes

\begin{displaymath}
\left| 
\begin{array}{ccc}
m_g &  z & \sqrt{2} z \\
z & m_{s\overline{s}} & 0 \\
\sqrt{2} z & 0 & m_{u\overline{u} + d\overline{d}}. 
\end{array} 
\right| 
\end{displaymath} 

Here $z$ is the annihilation amplitude for quark-antiquark into a
glueball, $m_g$ is the glueball mass before mixing with quarkonium, and
$m_{s\overline{s}}$ and $m_{u\overline{u} + d\overline{d}}$ are
quarkonium masses before mixing with the glueball and each other.  The
four unknowns in this matrix can be determined from four observed
masses. Our key assumption, described above, is that the isosinglet
$m_{u\overline{u} + d\overline{d}}$ before mixing is the same as the
observed isotriplet mass, 1450 MeV. In addition, we take the masses of
the physical, mixed states with the largest contributions from
$u\overline{u} + d\overline{d}$, $s\overline{s}$ and the glueball to
be, respectively, the observed resonance masses of 1390 MeV, 1500 MeV
and 1710 MeV.

Adjusting the parameters in the matrix to give the physical eigenvalues
we just specified, the unmixed masses and mixing parameter become
\begin{eqnarray}
m_g & = & 1635 \: {\rm MeV}, \nonumber \\
m_{s\overline{s}} & = & 1516 \: {\rm MeV}, \nonumber \\
z & = & 77 \: {\rm MeV}, \nonumber
\end{eqnarray}
while the physical mixed states as linear combinations
of the unmixed states are
\begin{eqnarray}
\lefteqn{| 1710 >  = } \nonumber \\
& & 0.87  | g >  + 0.34  | s\overline{s}>  
+  0.36  |u\overline{u} +d \overline{d}>, \nonumber  \\  
\lefteqn{ | 1500 >  = } \nonumber \\ 
& & -0.19  | g >  + 0.90  | s\overline{s}>  
-  0.40  | u\overline{u} +d \overline{d}>, \nonumber  \\ 
\lefteqn{| 1390 >  = } \nonumber \\
& &  -0.46  | g >  + 0.28  | s\overline{s}>  
+  0.84   | u\overline{u} +d \overline{d}>. \nonumber 
\end{eqnarray}
All mixed and unmixed states vectors here are normalized to 1.  Measured
in probability, the mixed glueball is more than 75\% pure glueball and
the mixed $f_0(1500)$ is more than 75\% pure $s\overline{s}$.  The mixed
$f_0(1390)$ is 71\% $u\overline{u} + d\overline{d}$ and 8\%
$s\overline{s}$ for a total of about 79\% pure quarkonium.

The negative sign of the contribution of $| u\overline{u} + d
\overline{d}>$ to $f_0(1500)$ can
lead to interference with the decay amplitude coming from $|
s\overline{s}>$ and suppress the width for $f_0(1500)$ decays
to $K\overline{K}$.
Assuming, for example, SU(3) symmetry of the coupling constants for
unmixed scalar quarkonium to $K\overline{K}$, we obtain a rate for the
mixed state $f_0(1500)$ which is about 40\% of the rate for 
unmixed $|s\overline{s}>$ decay to $K\overline{K}$.  
The negative sign of the $| g >$ coefficient will lead to a further
suppression if, as simple models suggest, $|s\overline{s}>$ decay to pairs
of pseudoscalars has the same sign as glueball decay to pairs of pseudoscalars.

On the other hand, since the glueball probability in $f_0(1390)$
is nearly six times that in $f_0(1500)$, we expect
the rate for the decay of $J/\Psi$ to $\gamma f_0(1390)$ to be
significantly larger than the rate for decay to $\gamma f_0(1500)$.
This expectation is supported by Mark III data~\cite{MarkIII}.

\end{document}